\documentclass[sigconf]{acmart}

\usepackage{wrapfig}

\AtBeginDocument{%
  \providecommand\BibTeX{{%
    \normalfont B\kern-0.5em{\scshape i\kern-0.25em b}\kern-0.8em\TeX}}}
    
\PassOptionsToPackage{numbers, sort, compress}{natbib}

\newcommand{\footnotetab}{\textcolor{white}{----}}

\copyrightyear{2021}
\acmYear{2021}
\setcopyright{acmlicensed}\acmConference[EAAMO '21]{Equity and Access in Algorithms, Mechanisms, and Optimization}{October 5--9, 2021}{--, NY, USA}
\acmBooktitle{Equity and Access in Algorithms, Mechanisms, and Optimization (EAAMO '21), October 5--9, 2021, --, NY, USA}
\acmPrice{15.00}
\acmDOI{10.1145/3465416.3483289}
\acmISBN{978-1-4503-8553-4/21/10}

\begin{document}

\title{Accuracy-Efficiency Trade-Offs and Accountability in Distributed ML Systems}


\author{A. Feder Cooper}
\affiliation{%
   \institution{Cornell University, Department of Computer Science}
   \country{USA}}
   \email{afc78@cornell.edu}

\author{Karen Levy}
\affiliation{%
   \institution{Cornell University, Department of Information Science \& Cornell Law School}
   \country{USA}}
  \email{karen.levy@cornell.edu}

 \author{Christopher De Sa}
 \affiliation{%
   \institution{Cornell University, Department of Computer Science}
   \country{USA}}
 \email{cmd353@cornell.edu}


\begin{abstract}
Trade-offs between accuracy and efficiency pervade law, public health, and other non-computing domains, which have developed policies to guide how to balance the two in conditions of uncertainty. While computer science also commonly studies accuracy-efficiency trade-offs, their policy implications remain poorly examined. Drawing on risk assessment practices in the US, we argue that, since examining these trade-offs has been useful for guiding governance in other domains, 
we need to similarly reckon with these trade-offs in governing computer systems. 
We focus our analysis on distributed machine learning systems.  Understanding the policy implications in this area is particularly urgent because such systems, which include autonomous vehicles, tend to be high-stakes and safety-critical. 
We 1) describe how the trade-off takes shape for these systems, 2) highlight gaps between existing US risk assessment standards and what these systems require to be properly assessed, and 3) make specific calls to action to facilitate accountability when hypothetical risks concerning the accuracy-efficiency trade-off become realized as accidents in the real world. We close by discussing how such accountability mechanisms encourage more just, transparent governance aligned with public values. 
\end{abstract}

\maketitle
\pagestyle{plain} 

\section{Introduction} \label{sec:intro}

Engineering is defined by trade-offs---by competing goals that need to be negotiated in order to meet system design requirements. One of the central trade-offs, 
particularly in computer science, is between \emph{accuracy} and \emph{efficiency}. There is an inherent tension between \emph{how correct} computations are and \emph{how long} it takes to compute them. 
While this trade-off is of general relevance, it plays out in various ways across  
computing: in computer hardware, circuits can use approximation techniques to relax constraints on accuracy---on how they perform bitwise computations---to speed up performance; in image processing, compressing pixels causes a loss in accuracy of the image being represented, but also furthers space-efficiency by requiring less memory for storage. In fact, such trade-offs are so abundant in computing 
that they have even given rise to its own subfield, \emph{approximate computing}~\cite{moreau2018taxonomy, mittal2016apsurvey}, which studies how different domains resolve the question of 
how much inaccuracy can safely be permitted for the sake of increased efficiency~\cite{sampson2015thesis}.


While the trade-off is commonly acknowledged in computer science, its policy implications remain poorly examined. We provide a starting point, in which we focus our analysis on \emph{distributed ML systems} using the running example of autonomous vehicles (AVs). We make this choice for two reasons. The first is urgency: AV development has made such significant strides that by 2040 at least 75\% of cars will have some level of autonomy
~\cite{newcomb2021avs}. Second, while AVs promise to improve overall driving safety,\footnote{The international effort to deploy AVs is motivated in large part due to AV technology's promise to increase automotive safety---that replacing human drivers with automated ones will protect millions of lives. Conservative estimates indicate that in 2035-2045, the decade in which AVs are targeted to reach widespread deployment, 585,000 lives will be saved worldwide~\cite{intel2017avs}.} they will also create new risks~\cite{ooida2020avs, boudette2021tesla}. As we show, some of these risks directly result from the accuracy-efficiency trade-off and the choices made to implement it~\cite{ntsb2019ubercrash}. In particular, the trade-off is tunable and context-dependent: It is not an all-or-nothing choice, and appropriate tuning depends on both a system's goals and deployment environment. 
Choices in different contexts will entail different emergent behaviors in technical systems---behaviors that are potentially high-stakes if, for example, they affect overall system safety.

We argue that the accuracy-efficiency trade-off exposes a high-level abstraction that policymakers should use to help hold such systems accountable.\footnote{We emphasize that this is \emph{not the only} such tool policymakers should have for holding these systems accountable. Other accountability mechanisms are also necessary, such as those that can assess hardware failures~\cite{abraham2019responsibility, surden2016avs, aaj2017avs}, the explainability of ML models~\cite{kroll2017aa}, and the impact of variance in automated decision-making~\cite{forde2021model}.} Rather than operating at one of two extremes---
solely having policymakers rely on technical experts to make high-stakes decisions or inundating policymakers with underlying low-level technical details---we advocate for something in between: Researchers should focus on providing correctness and performance guarantees, and should build 
tools to help policymakers reason about these guarantees. These tools should help expose the 
uncertainty in distributed ML systems. This 
would facilitate lawmakers' ability to assess whether trade-off implementations are aligned with safety goals, and to regulate the 
risk of deploying high-stakes systems like AVs. 
We emphasize \emph{distributed systems} because much of the sociotechnical conversation in ML has focused on \emph{algorithmic} fairness. This has left the systems components---notably, scalability, speed and their impact on correctness---under-explored in terms of their policy implications. 
As a result, ML \emph{systems} present under-examined challenges for technological accountability. We take the initial steps to bring some of these challenges to light, and suggest a novel framing for how to hold such systems accountable. This contribution demonstrates the need for mandatory risk assessment tools for  
distributed ML systems. We contend that, without such tools, effective public oversight of these systems will not be possible. Instead, we run the risk of manufacturers ignoring accountability mechanisms when constructing ML systems---or worse, deliberately making these systems difficult to assess in order to obscure responsibility when accidents occur. In both of these scenarios, the burden would fall on individual victims to prove manufacturer responsibility. This dynamic would make accountability quite difficult to achieve; 
the power and resource imbalances between individual victims and large ML-system manufacturers would make tort or other civil litigation infeasible~\cite{abraham2019responsibility}. \\ 

\noindent\textbf{Contribution.} Our analysis focuses on the US, but elicits principles that apply more broadly. We have chosen AVs as our central example because navigating the trade-off appropriately has already proven an urgent concern, notably in assessing Uber's 2018 AV crash~\cite{ntsb2019ubercrash}. To make our case, we survey relevant concepts and examples from law and computer science, and then synthesize this discussion to advocate for a concrete policy contribution, which we direct toward the National Highway Transportation Safety Authority (NHTSA).
\footnote{Approaching our topic in this interdisciplinary manner leads us to follow a nontraditional format. We need to justify our conceptual contribution in two directions, and thus provide a significant amount of relevant background information concerning how the accuracy-efficiency trade-off translates to both law and computer science.} We first discuss how the trade-off functions in relation to decision-making in disciplines other than computing, most notably in US risk assessment policy (Section \ref{sec:price}). Then, we provide an analogous discussion for ML algorithms and distributed ML systems (Section \ref{sec:computing}). 
We argue that reasoning about accuracy-efficiency trade-offs and accountability in highly technical domains is not a new problem. This suggests that, with the right technical tools, we can similarly hold high-stakes, distributed ML systems like AVs accountable (Section \ref{sec:policy}) with respect to how they implement analogous trade-offs. We close by discussing how such tools for increased accountability encourage more just, transparent governance aligned with public values (Section \ref{sec:conclusion}). 

\section{The Ubiquity of Accuracy-Efficiency Trade-Offs} \label{sec:price}

The trade-off at the heart of this paper is not unique to computing. It can be observed in a range of domains, many of which are regulated in the US, including law, the economy, and public health.\footnote{The accuracy-efficiency trade-off is also salient in other aspects of governance, including wartime intelligence gathering. The ``fog of war'' concerns the inherent tension between gathering more accurate intelligence about an opponent or enemy and acting on that intelligence before it becomes stale and loses its usefulness~\cite{clausewitz1832fog}.} In these disciplines, efficiency often can be thought of interchangeably with speed. For example, in decision theory, the time-value of information is an important concept for making choices. There is a cost to gathering increasingly accurate information: Waiting to act 
is itself an action---one that can have more negative consequences than acting earlier on imperfect information.\footnote{\citet{kahneman1982uncertainty} elaborate on this idea in their well-known cognitive psychology research concerning reasoning about uncertainty. They argue that humans use various heuristics to make decisions more efficiently, often acting on biases they have due to incomplete information. There is a tension between taking the time to gather more information and making a more informed decision---between the speed of making a decision and the quality of information used to make it.} \citet{sunstein2002heuristics} connects this idea to the potential hazards of using heuristics in legal decision-making. Nevertheless, he observes that heuristics are common (and necessary) 
to obtain a suitable balance between efficient resolution and the ``best'' (i.e., most accurate) adjudicative outcomes.\footnote{Due process is perhaps the most notable, encompassing example of balancing both values in US law.} 
For example, a number of rules in US civil and criminal procedure---speedy trial requirements, local filing deadlines, statutes of limitations---impose time constraints for the sake of efficient case resolution; these values must be balanced against needs for thorough fact-finding and argumentation. The standard for preliminary injunctive relief in the US requires courts to predict whether irreparable injury will occur because of the passage of time, if relief is not granted before the (often lengthy) full resolution of a case~\cite{lichtman2002injunctive}. Federal Rule of Evidence 403 allows for the exclusion of relevant evidence from a court proceeding if the probative value of that evidence is substantially outweighed by a danger of undue delay. 
These and other rules promoting judicial efficiency are, in the words of Justice Oliver Wendell Holmes, ``a concession to the shortness of life''~\cite{Reeve_Dennett}---they attempt to balance between the twin goals of getting matters right and getting them done, with recognition that there is real social value to each. 

Debates about the merits of the ``precautionary principle'' in policymaking also reflect the trade-off. The precautionary principle advises extreme caution around new innovations when there is substantial unknown risk; it places the burden of proof on risk-creating actors (like chemical plants) to provide sufficient evidence that they are \emph{not} producing significant risk of harm. As with speedy trials, there is a trade-off between the time it takes to gather evidence---to understand the risk landscape---and making informed decisions based on this landscape.\footnote{There are legal rationales on both sides of the spectrum with regard to how this trade-off should be implemented. For example, critics of the precautionary principle could be said to favor efficiency. They find the principle to be too stringent with regard to the burden it places on accuracy; it is ``literally paralyzing'' in its attempts to regulate risk~\cite{sunstein2003precaution}. On the other side, others argue that the precautionary principle provides a valuable way to reason about preventing harm by shifting the burden of proof of safety to potential risk creators. They are supportive of the fact that the principle requires actors to justify the risks they create: It is worth the time cost to gather information, such that it is possible to better manage risk in the context of scientific uncertainty~\cite{sachs2011precaution}.} A notable example of the precautionary principle demonstrating the trade-off in action concerns public health management of the SARS outbreak in the early 2000s. During the early outbreak of the disease, there was significant uncertainty around the risk of it spreading and how lethal it could be. The principle was adopted as a public health value at all of the disease epicenters: Individuals who were even remotely suspected of having come into contact with SARS were placed under strict quarantine. 
Years later, (pre-COVID-19) critics argued that mass quarantining led to a tremendous and unnecessary loss of liberty. They made this case based on analysis that indicated 66\% fewer individuals could have been quarantined with the same public health outcome (i.e., 
it would have still been possible to prevent a SARS pandemic)~\cite{chowkwanyun2016health}.\footnote{We are not yet at a time in which such retrospective analysis regarding the precautionary principle can be conducted for the ongoing COVID-19 pandemic. Nevertheless, the trade-off has still played a role in an additional public health context: antibody tests. The World Health Organization (WHO) has recently argued that, prior to certifying COVID-19 antibodies for treatment, it is necessary to \emph{guarantee} that such antibodies confer immunity to the virus. Several medical professionals have challenged this mandate from WHO, highlighting the time-sensitive nature of taking action in the pandemic: ``Demanding incontrovertible evidence may be appropriate in the rarefied world of scholarly scientific inquiry. But in the context of a raging pandemic, we simply do not have the luxury of holding decisions in abeyance until all the relevant evidence can be assembled. Failing to take action is itself an action that carries profound costs and health consequences.'' More generally, it is the norm for healthcare practitioners to act on incomplete information---to balance potential inaccuracies in available data with the urgency to treat serious conditions~\cite{weinstein2020covid}.} \\

\noindent\textbf{US federal risk assessment policy.} The examples above provide an intuition for how pervasive the accuracy-efficiency trade-off is in different domains, and how it is reasoned about to guide decision-making. Beyond this intuition, the trade-off is implicated more formally in US federal risk assessment standards and regulatory rule-making. Risk assessment policy acknowledges that, no matter how much time and resources one spends gathering scientific knowledge to assess risks, it will ultimately always be necessary to make decisions with uncertainty---to pass judgments in the face of incomplete information~\cite{nrc1983riskassessment, nrc1994riskassessment}.\footnote{As \citet{levy2016transparency} note, it is the epistemological nature of science itself that makes uncertainty inevitable in science-based policymaking: ``Agencies charged with protecting public health and the environment must make decisions in the face of scientific uncertainty, because science by its nature is incomplete and only rarely provides precise answers to the complex questions policymakers pose.''} There is always a degree of imprecision in scientific knowledge's ability to capture what is true, and that knowledge is constantly subject to revision in light of newly collected information. That is, taking more time to gather information can increase accuracy, but is directly at odds with efficiency in decision-making. 

In risk assessment, this trade-off is framed in terms of \emph{ex ante} (before-the-fact) and \emph{ex post} (after-the-fact) risk-mitigating interventions. The AI safety and fairness communities sometimes use the terms \emph{assessment} and \emph{audit}, respectively for \emph{ex ante} and \emph{ex post} \cite{falco2021audit}. \emph{Ex ante} mechanisms embody the precautionary approach: They emphasize collecting evidence about potential risks before approving a new substance or technology. For example, the US Food and Drug Administration (FDA) typically requires multiple phases of clinical trials before a new drug is approved for use (i.e., ``premarketing approval''~\cite{nrc1994riskassessment, nhtsa2016avs}). This \emph{ex ante} regulatory authority is deliberately slow for the sake of increased safety.\footnote{The FDA is empowered to require drug companies to submit sufficient data, such that a detailed risk assessment can be conducted before the drug goes on the market. This process can take a lot of time, and is not always conducted without criticism concerning choosing ``safety'' over ``efficiency''. For example, such critiques are common when swift approval has known safety benefits, but is delayed in favor of evaluating the presence of unknown (potentially non-existent) health risks. Debates concerning the FDA and this accuracy-efficiency trade-off have been particularly relevant recently concerning approving COVID vaccines for children~\cite{parkerpoper2021covid}.} In contrast, for efficiency, other agencies 
concentrate their authority in \emph{ex post} ``post hoc mechanisms''~\cite{nrc1994riskassessment}.\footnote{These mechanisms tend to require that agencies, rather than companies, acquire the data necessary to determine responsibility after an undesirable outcome occurs.} 
NHTSA has relatively weak \emph{ex ante} authority for determining what types of vehicles are safe to drive; its strongest authority is the ability to recall faulty cars \emph{ex post}~\cite{nhtsa2016avs,vinsel2019cars}.\footnote{NHTSA has the ability to set safety standards, and then verifies that manufacturers have met them through a self-certification process. In other words, manufacturers certify themselves as ``safe,'' rather than NHTSA soliciting data from manufacturers and performing the certification themselves~\cite{vinsel2019cars, nhtsa2016avs}.} NHTSA 
favors lack of \emph{ex ante} regulation as a way to ensure speedy development and deployment of new car technology, even if such lack of regulation comes with a cost in correctness in that technology. These are just two examples illustrating opposite choices concerning how accuracy and efficiency relate to \emph{ex ante} and \emph{ex post} enforcement. This trade-off spectrum applies to the risk assessment and rule-making practices of numerous other US agencies, including the Environmental Protection Agency (EPA), Occupational Safety and Health Administration (OSHA), and Consumer Product Safety Commission (CPSC), which each have different, domain-specific \emph{ex ante} and \emph{ex post} biases. 
Despite these differences, reports from the National Research Council (NRC) recognize that there are cross-cutting elements of risk assessment
~\cite{nrc1994riskassessment,nrc1983riskassessment}. The reports provide general recommendations for improving standards for accounting for uncertainty and its relationship to risk, such as 
clarifying the assumptions that inform model construction to elucidate model uncertainty. The NRC advocates for the importance of teasing out these low-level details, and communicating them to both decision-makers and the public, in order to ensure that policy goals reflect the known risk landscape. \\

\noindent\textbf{Toward policymaking for distributed ML systems.} 
This discussion shows that accuracy-efficiency trade-offs are a useful and natural way for policymakers to regulate varied, complex technical domains. We therefore ask: Why not use this framework for making policy concerning distributed ML systems? The specifics of the domain may vary---notably, real-time systems involve high speeds not present in, for example, evaluating the safety of new chemicals. Nevertheless, US risk assessment policy indicates that reasoning about accuracy-efficiency trade-offs, and their relationship to risk, is not a new problem. We therefore contend that reasoning about underlying accuracy-efficiency trade-offs can enable risk assessment and management for these emerging technologies. However, translating the above regulatory framing to this domain presents novel challenges. We will require new tools, which we clarify in Sections \ref{sec:computing} and \ref{sec:policy}, to reason effectively about similar trade-offs in distributed ML systems---tools that expose the particular type of uncertainty in real-time, distributed, automated decision-making. These tools will help us gather the data necessary for appropriate risk assessment and policymaking. 
Before we can describe these tools, we clarify that accuracy-efficiency trade-offs are an appropriate abstraction for accounting for the behavior of distributed ML systems. Having explained how reasoning about such trade-offs is useful for policymaking, we next make our case from a technical perspective. 


\section{Trading off Accuracy and Efficiency in Computing} \label{sec:computing}

\begin{figure}
  \begin{center}
    \includegraphics[width=0.45\textwidth]{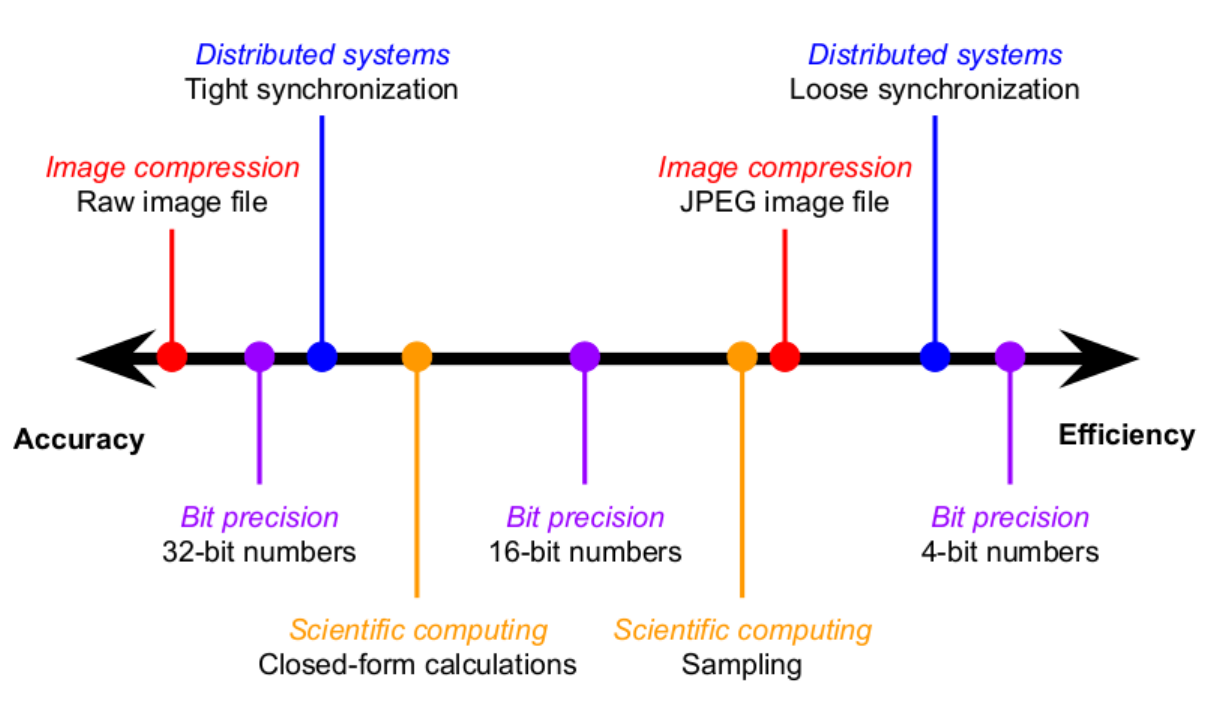}
    \vspace{-.35cm}
    \caption{Computing examples of the accuracy-efficiency trade-off spectrum: \textcolor{red}{Image compression}, \textcolor{violet}{bit precision}, \textcolor{blue}{distributed systems}, and \textcolor{orange}{scientific computing}.} 
	\label{fig:tradeoff}
  \end{center}
  \vspace{-.62cm}
\end{figure}

Accuracy-efficiency trade-offs are particularly relevant across computing.\footnote{The accuracy-efficiency trade-off is arguably a central concern for the entire field of computing. \citet{ohm70inefficiency} call efficiency the ``cardinal virtue'' of computing in order to discuss what they view as exceptional cases of inserting inefficiency into computer systems---what they term ``desirable inefficiency.'' Instead, viewing the accuracy-efficiency \emph{trade-off} as central enables us to not refer to ``inefficient'' computing models (e.g. cryptography) as exceptional. We conceive of them as implementing the trade-off at one end of the accuracy-efficiency spectrum (with cryptography privileging accuracy), which strikes us as a more precise and generalizable statement.} 
To understand this, consider a familiar example---JPEG compression. Raw images tend to be very high resolution: They contain many, varied pixels per inch, and therefore require a lot of storage space. However, 
a compressed, JPEG version 
often suffices for high quality; combining neighboring pixels often is not detectable to the human eye. A JPEG also takes up less storage space and can lead to faster processing when doing photo editing since there are fewer pixels to consider; it is more space- and time-efficient. Reducing the accuracy of the image can lead to greater computational efficiencies. 

This type of trade-off spectrum 
forms the basis of \emph{approximate computing} (Figure \ref{fig:tradeoff}), which studies how 
a computer system can achieve certain performance benefits if it exerts less  
effort to compute perfectly accurate answers. In other words, it is possible to \emph{relax} accuracy in order to yield efficiency improvements~\cite{moreau2018taxonomy, mittal2016apsurvey, sampson2015thesis}.\footnote{We do not include the pathological case in which \emph{all} accuracy is sacrificed in order to do something really fast but completely wrong. Nevertheless, there are cases where an implementation could, for example, be wrong 40\% of the time (for increased speed) and still achieve certain application-specific quality goals.} As with JPEGs, relaxing the accuracy does not necessarily have negative consequences; rather, it is possible that decreased accuracy has no observable impact for a particular application. That is, some applications are tolerant of inaccuracy; they are error resilient. 
Similar to non-computing domains, tools for reasoning about the trade-off inform decisions about how to implement it. 

Computer scientists create theoretical tools to characterize the trade-off, which they leverage to determine 
the right implementation in different applications. 
Formal reasoning about the trade-off can yield application-specific quality metrics, where quality can be thought of as whether a program produces ``good enough'' results. Often, ``good enough'' cannot be guaranteed with complete certainty, but can be verified with high probability. Leaving room for uncertainty allows for edge case behaviors that fall below the specified quality threshold. Quality metrics therefore capture how much an approximation is allowed to deviate from the precise version's results. Computer scientists can then design software that requires a certain degree of program quality with a certain (high) probability~\cite{sampson2015thesis}.\footnote{A popular example of this comes from Amazon's cloud computing services (AWS). Their cloud storage service provides ``11 9's'' of reliability with regard to storing data objects, meaning that 99.999999999\% of the time saving such objects to the cloud occurs without error~\cite{amazon2020s3}.} \\

\noindent\textbf{Accuracy-efficiency trade-offs in ML. $\:$} Such trade-offs are a salient concern across ML. Notably, in deep learning, there is an ongoing, increasing emphasis on training larger models to yield more accurate results. This comes with host of efficiency challenges, including significantly increased training time, model storage requirements, and energy usage~\cite{kaplan2020scaling}.\footnote{The trade-off notably did not first become relevant with (though is arguably increasingly urgent due to) the advent of modern statistical ML. Several influential papers on artificial intelligence (AI) from the 1980s and 1990s also demonstrate the potentially high impact of appropriately dealing with accuracy-efficiency trade-offs~\cite{horvitz1987constraints, boddy1994deliberation}.}
Moreover, 
ML models perform inference that is not always correct; to be robust, models need to tolerate a certain degree of inaccuracy. 
This notion of error resilience (or inaccuracy tolerance) varies for different ML algorithms. Regardless of particular differences, there is a general tension between \emph{correctness} and \emph{performance}.\footnote{For example, the correctness of a training algorithm can be understood as whether or not the algorithm converged to the distribution we set out to learn, i.e., \emph{Did we learn the right model?} Its performance indicates whether convergence to the distribution---whether correct or incorrect---happened in a timely manner, i.e., \emph{How fast did we learn the model?}} In fact, relaxing accuracy to increase efficiency is a requirement in many learning domains. Otherwise, computations can be so slow to perform that they become intractable. One relaxation strategy\footnote{These examples are far from exhaustive. We picked these two because they reflect commonly-used strategies across various ML areas, rather than niche techniques relevant to only one specific subfield. A third such example is resource-constrained techniques, which involve smaller computers, such as Internet of Things (IoT) devices and sensors. With the advent of IoT in recent years, there has been a significant increase in the variety of computers available and a corresponding increase in the variety of computations we wish to run on them. For example, an Amazon Echo serves up answers to spoken language questions; however, it also has limited on-board capabilities to perform computations locally. These limitations take several forms. For example, such devices might not have a lot of power to process data quickly or might lack storage capacity for large amounts of data. As a result, such devices often only have smaller, coarser-grained models in local memory, which can be used for quickly returning (potentially less accurate) inference results. Often, these devices can communicate with more sophisticated computers over the Internet, offloading computation or storage to those computers. Because these computers have more memory and processing capabilities, they can store larger models that are capable of more nuanced inference. \\ \footnotetab However, this communication exposes another accuracy-efficiency trade-off; it takes time to send the data to a remote computer, perform some (more accurate) computation, and then return a response to the device~\cite{birman2019cloud}. That computation may be more accurate due to using a larger, finer-grained model, but achieving that accuracy comes with a cost in speed. Conversely, doing the computation locally on the device would be faster; however, due to the device's more limited computational resources, it will not necessarily be as accurate. For example, prior work in computer vision considers how to handle the trade-off when performing ML on mobile devices, such as smart phones~\cite{howard2017mobilenets}. This work uses manually-tunable parameters that allow the model developer to strike the right balance for particular learning problems. Depending on the application domain, a model developer can tune a larger model that uses more resources (i.e., a model that is slower or uses more memory but is more accurate) or one that is smaller and uses fewer resources (i.e., a model that is faster or uses less memory but is less accurate). Aside from being faster, there are several reasons why local computation and storage might be desirable for a mobile application, as opposed to offloading these requirements to more powerful remote computers. Notably, local computation can ensure privacy, as the learned model and collected data never leave the mobile device~\cite{wang2019privacy}. \\ \footnotetab A fourth such example of a strategy is low-precision computing, or quantization, to use fewer bits to speed up computation (i.e., decrease accuracy for increased scalability)~\cite{desa2017async, gong2014quantize, courbariaux2015binaryconnect, alistarh2017qsgd, gupta2015lowprecision,han2015deep}. This method, sometimes called quantization, is similar to the idea of floating-point precision---how much accuracy the computer can capture based on how many bits it uses to represent numbers (Figure \ref{fig:tradeoff}. Computing with more precise floating-point numbers is more computationally expensive; it tends to take more time and memory (i.e., sacrifices efficiency) but can capture a more accurate range of results. Much work in machine learning explores using low-precision numbers to achieve faster results. This work relaxes requirements on the accuracy of the trained model in order to achieve these speed-ups. There is also a spectrum at play here. It is possible to vary the number of bits of precision: More bits yield higher accuracy and slowdowns, while fewer bits require less time per computation and thus potentially sacrifice some correctness. Depending on a particular application's tolerance to error, this sacrifice in accuracy can be worth the speed-ups it creates~\cite{desa2018halp}. It is also possible to implement low-precision computing in hardware~\cite{carmichael2019dnn, colangelo2018fpga, zhao2019overwrite}. \\ \footnotetab In general, we must also consider how the hardware specifications of the computer running the algorithm might also impact that behavior. Surely this is important, as different computers have different computing capabilities due to varying hardware; a NASA supercomputer has more computational resources than a personal laptop. As with the subsampling, a low-bit-precision sacrifice in accuracy does not necessarily require sacrificing overall correctness, if in expectation the algorithm can still theoretically guarantee learning the right distribution. \\ \footnotetab Notable examples of subfields with specific trade-offs include reinforcement learning (RL) and Markov chain Monte Carlo (MCMC). In RL, there is the well-known exploration-exploitation trade-off (more exploration increases accuracy and more exploitation increases efficiency)  \cite{ishii2002rl, jin2020rl}. In MCMC, algorithms exhibit scalability-reliability trade-offs (scalability corresponds to efficiency, reliability to accuracy) \cite{zhang2020tunamh}.} is \emph{subsampling} during training, which involves using a subset of the dataset in place of the entire dataset to compute model updates faster.\footnote{Performance directly relates to the size of the task on which we conduct learning. Intuitively, if a learning algorithm is slow on tasks with small datasets, then that algorithm will be slow, if not computationally intractable, on much larger ones. This relationship between runtime and task size often exists due to coupling between the computation done by the learning procedure's optimization algorithm and the task's dataset size. For example, when computing the gradient needed to determine which direction the learning algorithm should step for its next iteration, it is often necessary to sum over every data point in the dataset.} Even though each iteration is less accurate (but more efficient), some algorithms can still guarantee overall high-quality (i.e., statistically correct) results.\footnote{A very common approach for improving efficiency is to use a subsample or \emph{minibatch} of the dataset, rather than the whole dataset, when performing calculations. In the case of computing gradients, instead of using a \emph{full batch} (i.e., the whole dataset) we use a randomly sampled subset of the data points, which involves spending less time on the computation of a particular iteration. Stochastic Gradient Descent (SGD) is an example of an algorithm that takes this approach, in which using a minibatch can  have minimal impact on the overall accuracy of the learned model. A particular iteration of the algorithm will have less accuracy when computing the gradient; but, when run for lots of iterations, the final result is usually still statistically correct. In expectation, we can learn the same distribution as if we had been using the whole dataset in each iteration; we can often theoretically guarantee robustness~\cite{bottou2018sgd}. Moreover, the decision to subsample is not all-or-nothing; it is a spectrum. It is possible to vary the minibatch size the algorithm uses. Larger minibatches---especially those that approach the size of the full dataset---require more time but are also more accurate per iteration. Conversely, smaller batch sizes make each iteration faster and more scalable to larger datasets, but in doing so sacrifice accuracy per iteration. Determining the right sweet spot in this trade-off often depends on the particular learning task, and often falls under the area of study called hyperparameter optimization~\cite{feurer2019optimization}.} \emph{Asynchrony} enables 
different computer processes or threads\footnote{Threads and processes are mechanisms for parallelization within a computer \cite{arpacidusseau2018os}. A process can have multiple threads running at the same time. For example, this is what allows a text editor (which is running in a process) to simultaneously enable displaying both typing and syntax-error highlighting in real-time. Each of these functions happens in its own thread, within the process of running the text editor application.} to
perform computations side-by-side and combine the results.\footnote{In other words, asynchrony can speed up ML since multiple parts of the learning problem can be computed at once.} 
This is more efficient but, depending on how the results are combined, can also lead to decreases in accuracy: If different processes work on overlapping parts of the overarching computation, one process can potentially overwrite the value recorded by the other out of sequence~\cite{desa2017async, alistarh2018convergence, lian2017asynchronous, Niu2011hogwild}. This can be avoided by forcing processes to coordinate their updates, but such coordination takes time; it increases accuracy, but decreases efficiency.\footnote{Out-of-sequence overwriting from asynchrony can be worth the speed-ups it enables; it is still possible---though not always guaranteed---to compute good quality learning estimates~\cite{desa2016gibbs}. Moreover, asynchrony can be used in conjunction with minibatching or resource-constrained devices, yielding additional accuracy-efficiency trade-offs.} \\ 

\noindent\textbf{Implications in real-world ML systems.} 
We have 
provided examples of the trade-off in ML \emph{algorithms}, 
but have not yet considered how the trade-off behaves in \emph{deployed systems}---systems that consist of multiple computers that 
work together to solve large, complex problems.\footnote{Such systems often introduce additional asynchrony: Instead of one computer running an algorithm to solve a task, multiple computers work together in parallel.} 
Our 
aim is to understand the particular trade-off challenges in such \emph{distributed ML systems}, 
so we need to account for the ``distributed systems'' component just as much as ``ML''. The distributed setting is what enables potentially life-saving technology like AVs.\footnote{These systems reflect a triumph of new systems abstractions, not just innovations in ML~\cite{birman2019cloud}.} Importantly, new risks emerge when such fast, scalable systems are deployed in the real world. For example, researchers recently built a model that they showed could outperform humans in identifying gay individuals using facial recognition technology~\cite{wang2018gay}.\footnote{This claim has been challenged by several researchers, notably \citet{leuner2019replication}.} This disturbing result yielded a blizzard of media attention~\cite{hawkins2017gay, murphy2017gay}, yet it was also small-scale and slow. Consider a similar model, but one that is scalable and fast---integrated with a CCTV surveillance system serving real-time inference and deployed in a country hostile to LGBTQ rights. This may sound like science fiction, but low-latency, distributed vision systems already exist~\cite{wang2018china}. While this example is generative concerning the range of potential risks from 
ML systems, we focus on the risks related to accuracy-efficiency trade-off implementations.\footnote{As we note in Section \ref{sec:intro}, while we focus our discussion of the policy implications of accuracy-efficiency trade-offs in distributed ML systems, reasoning about such trade-offs in other parts of computing could also serve useful to tech policymaking. Similarly, we focus our analysis concerning accountability mechanisms to the accuracy-efficiency trade-off, even though distributed ML systems raise a variety of other accountability concerns, aside from this trade-off.} We next clarify how the trade-off is implicated in distributed computing, and then combine this with our ML discussion 
to show how the different tensions interact with each other. Considered together, 
ML and distributed computing trade-offs present especially challenging problems for real-time, high-impact systems like AVs. In Section \ref{sec:policy} we will ultimately argue that clarifying the relationship between these risks and trade-off choices can help policymakers hold such systems accountable. \\


\noindent\textbf{Accuracy-efficiency trade-offs in distributed computing.} In contrast to a single computer, a \emph{distributed system} is a network of computers that 
can work together to solve problems. Each computer has its own data and performs its own computations, and it shares data and computation results with other computers in the network when necessary. 
Because the computers are in distributed locations---whether in the same data center or across the world---
there are important considerations with regard to how efficiently information can be shared between them. When a computer contacts another in the system to request data, it takes time to complete the request and receive the data, reducing time-efficiency. There are also issues of accuracy between computers. Each computer has its own data---its own view of 
the state of the overarching system. That information is not complete: It is just a subset, which can conflict with the views of the other computers in the system.  
In other words, in distributed systems we can more specifically frame the accuracy-efficiency trade-off as a tension between \emph{consistency} and \emph{latency}\footnote{Latency can be informally thought of as the speed with which the system updates.}. There is a trade-off between all of the computers in the system having the same understanding of the data in the system and the time it takes to propagate that understanding throughout the system~\cite{abadi2012tradeoff, brewer12computer}. In distributed systems that update their data frequently it is quite difficult to quickly build a consistent, holistic understanding of the environment across different computers in the network.\footnote{One could informally view consistency is a moving target; each computer processes information locally faster than it can share it with the entire network.} Since it takes time to communicate, it is hard for computers to stay completely up to date with each other. For the sake of efficiency, individual computers in the system often need to make decisions in the presence of inconsistency.\footnote{Waiting for complete consistency across computers before an individual computer could make local changes would bring the entire system to a standstill. This is especially relevant if a computer in the system experiences a fault; to achieve strong consistency, before proceeding with local computation, all of the other computers would be waiting to hear from a computer that can no longer communicate with them (i.e., they could end up waiting indefinitely).} 

Particular distributed system implementations need to answer the question of how much application-dependent inconsistency and slowness they can each tolerate. 
To understand this spectrum, we will use the example of a social media website, which has computers hosting its data all over the world. A user   
tends to access the geographically closest computer server hosting the site; different users across the world therefore access different computer servers. Such a system favors efficiency (i.e., low latency) over the different computer servers being consistent with each other. It is more important to return the website to each user quickly than it is to make sure that every user is accessing the website with exactly the same data. This is one reason why on some social media sites it is possible to see out-of-order comments on a feed. To resolve its current state, the site aggregates information from across the system. It attempts to build a consistent picture, but limits how much time it spends doing so---sacrificing consistency---so that it can remain fast~\cite{decandia2007dynamo, lu2015existential, vogels2009eventualconsistency}. The system implements this choice via its communication strategy. Rather than contacting every computer in the system to construct a consistent picture, 
a particular computer only communicates with a subset. It trades off the accuracy it would get from communicating with every computer for the efficiency of communicating with fewer computers~\cite{hellerstein2019calm}.
Based on communication strategy, it is possible to quantify consistency and to measure it 
throughout a distributed system~\cite{lu2015existential, Shang2018rushmon}. Developers can reason about the degree of inconsistency their particular system can tolerate safely, and can detect and tune the system 
accordingly to also enforce an upper bound on latency~\cite{golab2011funandprofit,barbara1990controlledinconsistency, yu2000numerical}. \\

\noindent\textbf{Distributed ML systems: AVs as a case study.} 
We can now specifically consider accuracy-efficiency trade-offs in real-time (i.e., latency-critical) distributed ML systems. 
We will focus on AVs as a concrete example, 
which will facilitate making concrete policy recommendations (Section \ref{sec:policy}). An AV can be thought of as a distributed system of sensors.\footnote{This setting is further complicated by the fact that numerous vehicles can also be networked together (Vehicle-to-Vehicle, or V2V) and with other devices like smart traffic lights (Vehicle-to-Infrastructure, or V2I), which increase both the size and complexity of the system under analysis \cite{nhtsa2016avs, surden2016avs, transportation2014communications, fmvss2017communications}.} While each AV maintains its own local notion of the state of the environment, information that other AVs possess could also prove useful. If an accident is up ahead, an AV closer to the crash can communicate that information to those behind it, which in turn can apply  
their brakes and potentially prevent a pile-up. In such real-time transportation domains, accuracy and efficiency are both critical. Some ML applications may be able to tolerate wide margins of error, but in safety-critical domains a high degree of inaccuracy may be unsafe. The same goes for efficiency; such systems will need to make decisions quickly and, like the non-computing examples in Section \ref{sec:price}, there is an inherent trade-off between waiting to make a completely informed decision and making a decision fast enough for it to be useful~\cite{abadi2012tradeoff, brewer12computer}. 
What is unique here for AVs 
is the degree of time-efficiency needed. In some cases, inference decisions will be necessary at sub-second speeds, and will therefore be computed using inconsistent or uncertain information. This presents a challenge; in the face of this uncertainty, we need systems like AVs to be guaranteed (at least with very high probability) to be accurate. 
The urgency of resolving this problem is not merely a hypothetical situation; the accuracy-efficiency trade-off in fact played a crucial role in the Uber AV crash in 2018~\cite{ntsb2019ubercrash}, which we will return to in Section \ref{sec:policy}.

It is not entirely clear what the right trade-off implementation is for real-time systems like AVs~\cite{dietterich2018robustAI}. Unlike the example 
trade-offs in Figure \ref{fig:tradeoff}, AVs are mobile and deployed in varying environments. While those examples each indicate a single, static, application-dependent trade-off implementation, an AV might instead need to support a range of trade-offs given the dynamic nature of the environment. A particular trade-off implementation may need to depend on different operational design domains (ODDs) that vary by roadway type, geography, speed range, and lighting, weather, and other environmental conditions \cite{nhtsa2016avs, sae2021avs}. Some ODDs will be more efficiency-critical: It would be catastrophic for a car to take an extra half-second to be certain that there is a pedestrian directly in front of it~\cite{ntsb2019ubercrash}. In other cases, having an accurate sense of the environment may be more important than speed. 
For example, when detecting a deep pothole up ahead, it could be safer for a car to slow down to decide its course of action---to accurately determine if the hole is shallow enough for the car to continue on its course or deep enough to warrant veering off the road to avoid it.

As this example indicates, distributed ML systems raise different accuracy-efficiency questions than either distributed systems that do not involve ML, or ML systems that are not distributed. Since ML models (necessarily) approximate 
the world, it is possible for them to operate on data that are not completely accurate and still yield results that are correct \emph{enough}---that fall within the same bounds of imperfection that we deem tolerable. 
We can extend such inaccuracies beyond things like subsampling to include the data staleness 
inherent in distributed settings~\cite{bailis2012pbs, decandia2007dynamo,yu2000conits}.\footnote{Staleness is not the only property that can be tolerated; another example is numerical error that comes from asynchrony \cite{yu2000numerical}, which we elide for brevity.} Allowing for staleness 
increases efficiency, as the system does not need to wait to synchronize state  
before proceeding with its computation. As with a single computer, the overall output still \emph{can be} correct even when operating on 
stale data in a distributed setting; however, existing work in this field does not  guarantee such output \emph{must be} correct~\cite{alistarh2018convergence, gong2014quantize, lian2017asynchronous, Niu2011hogwild, desa2015taming, zhang2015sgdstaleness}. For AVs, this does not suffice; we want to be able to guarantee correctness\footnote{Of course, with those guarantees predicated by certain assumptions. At the very least, we need to bound the likelihood of incorrectness.} in order to be assured of safety. 

Such assurance will require us to reason differently about the behavior of distributed ML systems. Prior work has examined the trade-off at a high level by looking at correctness and speed metrics of end-to-end ML systems ~\cite{abadi2016tf, ho2013SSPParameterServer, li2014parameterserver, kosaian2019paritymodels, pan2016cyclades}; this work uses overall empirical performance results to tune the staleness of the underlying data storage layer. There is a fundamental mismatch in this approach: High-level performance metrics are used to \emph{indirectly} tune low-level system behavior (to, in turn, affect high-level performance), without formalizing the relationship between the two. This is an inversion of what we ideally would like to do: to formally evaluate the underlying accuracy-efficiency trade-off, and use this information to \emph{directly} tune distributed ML system behavior. As a result of this mismatch, tuning has generally been manually curated to the particular problem or absent, leaving an engineer to pick from predefined settings that enforce high accuracy guarantees over efficiency, ignore accuracy guarantees altogether in favor of efficiency, or attempt some middle-ground. While there is a valid spectrum of trade-off points, 
current large-scale ML systems tend to opt for efficiency over accuracy.\footnote{They focus on minimizing communication between computers in the system in order to be fast enough to scale to larger problems. Some of these systems can achieve orders of magnitude in efficiency improvements by dropping data updates without simultaneously destroying correctness~\cite{Niu2011hogwild, tsitsiklis1986stochastic}.} It is not clear these approaches will be safe for systems like AVs.\footnote{It may not always be safe for these systems to lose updates. Existing approaches to mitigate such losses in ML systems involve increasing communication between computers in the system. However, this strategy impacts the other side of the trade-off, leading to inefficiencies from bottlenecks in coordination between computers. This problem is similar to what exists in weakly consistent storage systems, which have side-stepped this issue by using semantic information to coordinate ``only when necessary''~\cite{dipippo1997semanticcc, molina1983semantic, weihl1988commutativity}.} It remains an open research question how safety-critical, real-time distributed ML systems like AVs should implement the trade-off.

\section{Accuracy-Efficiency Trade-Offs as a Mechanism for Accountability} \label{sec:policy}

Systems like AVs are really complex, but complexity should not serve as a rationale to preclude their regulation. Rather, the fact that these challenges remain unresolved presents an opportunity: Stakeholders aside from engineers can help shape implementations; they can inform accuracy-efficiency trade-off choices so that they align with the public's interests, 
not just those of manufacturers. This is why we have taken considerable space to clarify a variety of accuracy-efficiency trade-offs---from how they impact computing broadly to how they describe a range of possible behaviors for distributed ML systems. Though much of our prior discussion is well-acknowledged in technical communities (albeit, in other forms), to date the trade-off's implications have not been made legible to policymakers. The trade-off is not binary; it is a spectrum and can be treated like a tunable dial set appropriately to the context (Section \ref{sec:computing}). Our hope is that exposing this dial for distributed ML systems will provide a degree of technical transparency to lawmakers, such that high-stakes systems like AVs are not deployed without sufficient public oversight. We believe that explicitly exposing this trade-off provides a mechanism for holding these systems accountable for some of the risks they create. 

To do so, we address the gaps between existing risk assessment tools and what is needed to analyze accuracy-efficiency trade-offs in AVs. 
When an 
undesirable outcome occurs, we can examine accountability along two dimensions: the time window around the outcome, which we consider in \emph{ex ante} and \emph{ex post} divisions, and the actors that assess the system's behavior, which consist of computer scientists and policymakers. There is a region of overlap in which computer scientists can assist policymakers with \emph{ex post} evaluation and policymakers can frame \emph{ex ante} risks prior to deploying systems. We therefore propose a twofold call-to-action for enabling risk assessment in this domain: 1) Computer scientists must build tools to expose underlying accuracy-efficiency trade-offs and 2) Policymakers should use these tools to assess trade-off implementations, and meaningfully intervene to ensure implementations align with public values. We discuss these calls-to-action in terms of \emph{ex ante} and \emph{ex post} risk assessment gaps. \\

\noindent\textbf{Addressing \emph{ex ante} risk assessment gaps.} 
A system's ability to be assessed with respect to the accuracy-efficiency trade-off should be considered as important as every other technical feature. We therefore call on computer scientists to engage in research to build tools in ML systems that make their accuracy-efficiency trade-offs assessable. We explain what we mean by ``assessable'' via example and then suggest research directions to help make assessments possible. \\

\noindent\textbf{The 2018 Uber AV crash.} The 2018 Uber AV crash illustrates the importance of tools to assess the trade-off~\cite{ntsb2019ubercrash}. The crash resulted from the coincidence of several issues,\footnote{Together, the NTSB report generally summarizes these issues as reflective of a ``lax engineering culture'' around safety at Uber.} one of which had the accuracy-efficiency trade-off as its central problem. The AV remained inconsistent and indecisive for over 6 seconds.\footnote{The AV clearly had not implemented a robust inconsistency resolution strategy, as it this is a significant amount of time for a computer to not to resolve inconsistency.} 
By the time the sensors agreed about the presence of a pedestrian, the AV had already collided with her.\footnote{This example is far more complex than what we have glossed here. For example, there were no other cars on the road, so it seems certain that slowing down to take the extra time to resolve inconsistency would have been safe. Additionally, there was a human back-up driver; however, she was not paying attention. Even if she had been, it is not clear that she could have responded appropriately within 6 seconds, as average time for human take-over from an AV is 17 seconds \cite{nhtsa2015humanfactors}.} 
While the NTSB report is clear that the AV's sensors were inconsistent, it is not clear \emph{why} the AV could not make a decision. In this case, a granular explanation was not necessary to determine accountability, as 6 seconds is a very long time to be inconsistent. This AV was neither accurate nor efficient, indicating a sub-optimal trade-off implementation, as opposed to a well-reasoned choice, that led to a tragic outcome. In instances that are not as clear-cut, such as those that involve much tighter time windows, tools that provide granular explanations will be necessary to determine the difference between bugs and deliberate trade-off choices. \\

\noindent\textbf{Research directions for trade-off assessment tools.} We need novel trade-off assessment tools to evaluate more difficult cases. Such tools could help avoid certain risks, guaranteeing \emph{ex ante} specific desirable system behaviors while foreclosing the possibility of other undesirable ones. That is, in some scenarios it may be possible to reduce the tension between accuracy and efficiency by taking coordination between computers off of the critical path; this would enable greater computational efficiencies without sacrificing accuracy in those contexts~\cite{hellerstein2019calm}. For example, 
program analysis could help formally categorize underlying accuracy-efficiency trade-offs, and therefore facilitate building asynchronous systems with more effective concurrency control and theoretically provable correctness guarantees~\cite{roy2015homeostasis, molina1983semantic}.  
This would solve the mismatch in current asychronous ML: Instead of using high-level empirical observations to do ad-hoc, low-level system tuning (Section \ref{sec:computing}), we could directly tune the underlying trade-off to guarantee end-to-end performance behavior.
\footnote{More specifically, we could use program analysis to leverage the underlying semantics of the program and data to avoid synchronization (i.e., inefficiency); these techniques would enable performing efficient, provably correct asynchronous computation.} If program analysis indicates that strong consistency is not possible, we could weaken this requirement by instead bounding how much inconsistency is tolerable. We could perhaps even bound inconsistency such that the overall correctness of the asynchronous computation is not too severely impacted~\cite{,dipippo1997semanticcc, yu2000conits, yu2000numerical}. To make this idea concrete, consider that 
not \emph{all} of the AVs in the system will always need to communicate with each other. Instead, it will likely be sufficient for AVs to only communicate with others in an environment-dependent radius. Reducing communication to that radius would increase efficiency without decreasing accuracy, as AVs outside the radius would be too far away to have relevant information to communicate.\footnote{In other words, inconsistency between cars that do not need to communicate with each other is tolerable. We instead prioritize (limited) communication between relevant cars, where relevance is determined via automated reasoning about the underlying semantics of the problem. This example is extremely high-level---described at the level of individual AVs---for the purpose of clarity. Semantic analysis will expose lower-level (i.e., at the level of particular data points), less-intuitively-explainable opportunities for better concurrency control.}  

By providing such clear mechanisms to reason about accuracy-efficiency trade-offs, computer scientists expose a particular kind of decisional uncertainty that depends on time~\cite{horvitz1987constraints, boddy1994deliberation}. Clarifying this uncertainty does not, however, identify specific risks that automated decisions can bring about. 
Rather, it is up to policymakers to frame potential risks and to identify the normative, domain-specific values at play~\cite{jasanoff2016ethics, friedman2019values, flanagan2014values, goldenfein2020handoff}. 
Based on the uncertainty that computer scientists expose, 
policymakers should endeavor to assess \emph{ex ante} how much of the resulting risk is tolerable. Such \emph{ex ante} interventions could help narrow the space of potentially deviant system behavior, which in turn could help narrow the number of incidents examined \emph{ex post}. These interventions, though unlikely to be comprehensive, should clarify many of the risks in deploying these systems. 
However, it will not always be possible to preemptively fully analyze the risk landscape due to the amount of uncertainty in the system~\cite{sunstein2003precaution, smith2015opportunism}. Incomplete risk analyses will not necessarily prevent the deployment of real-time ML systems in practice; instead, policymakers will need to evaluate system behavior \emph{ex post}, after undesirable outcomes occur. A bad outcome will either reveal a risk that policymakers previously did not consider, with which they now need to contend, or it will implicate an acknowledged risk previously deemed acceptable. \\

\noindent\textbf{Addressing \emph{ex post} risk assessment gaps.} 
When deployed for long enough, high-stakes ML systems are likely to incur severe harms that we likely did not anticipate~\cite{perrow1999risk, vaughan1996challenger, nissenbaum1996accountability, smith2015opportunism}. This is where tools that expose the accuracy-efficiency trade-off, described above, can facilitate accountability after-the-fact: 
They could facilitate determining if a system has deviated further than expected from normal behavior (i.e., what \emph{ex ante} risk assessment deems to be acceptable)~\cite{sampson2015thesis}.\footnote{\emph{Ex ante} audit systems abound in security-related literature. For example, see ~\citet{lampson2004security, haeberlen2007peerreview, falco2021audit}.} In these cases, policymakers would still be able to hold the appropriate stakeholders accountable \emph{ex post}. 
We do not claim that policymakers need to understand 
low-level technical details to provide this oversight (e.g., the particulars of concurrency control algorithms). 
Rather, we are suggesting that surfacing higher-level trade-offs (that lower-level technical decisions entail) clarifies valid sites for potential policy intervention. 
Such trade-offs are the right level of abstraction with which policymakers can engage in order to reason about relevant policy goals; 
the accuracy-efficiency trade-off can clarify how lower-level engineering decisions relate to overall notions of system safety~\cite{sampson2015thesis}.

It is this reasoning that informs our second call-to-action: Policymakers should view the accuracy-efficiency trade-off as a regulable decision point at which they can meaningfully intervene. They already do so in other complex technical domains, for which they reason about risk and interventions (Section \ref{sec:price}). This suggests that, with the right tools integrated with distributed ML systems---like those we suggest above---
policymakers should also be able to do so for these systems. We do not articulate specific policies, as these will depend on a more comprehensive study of AV technology beyond the scope of this paper. Instead, we have used AVs as a guiding example to illuminate abstract technical concepts and their import for technology policy concerning accountability. 
It is possible to view this contribution is as an extension of existing risk assessment tools in computing. Contemporary policy debates about high-stakes ML applications in policing, 
transportation, and public health also involve concerns about what degree of accuracy we ought to demand from automated systems. These concerns often arise in 
attempting to minimize disparate outcomes across groups.\footnote{E.g., differential accuracy rates for face recognition along dimensions of race and gender~\cite{buolamwini2018gender, cooper2021emergent}.} But we contend that debates about the harms of inaccuracy are incomplete if they fail to 
reckon with the accuracy-efficiency trade-off. 
For policymakers, these debates will 
require 
trade-off assessment tools 
to analyze gaps between the expected risks and the actual behavior of distributed ML systems. 
For example, we could fairly pose to policymakers questions like the following: At what point is information sufficiently high quality to justify a system executing 
high-impact decisions? 
When is it safe for a system to spend more time computing decisions, particularly when more efficient heuristics do not sufficiently remove uncertainty? 
These tools will therefore take 
a step toward closing 
the ``responsibility gap''~\cite{jasanoff2016ethics}: Policymakers will have a more complete understanding of technology and will be better equipped to gauge the range of possibilities for its governance. This way, when technological failures occur, 
policymakers can \emph{ex post} more actively participate in the evaluation of how uncertainty in distributed ML systems contributes to risk.

\section{Conclusion: Toward More Just, Transparent Public Governance} \label{sec:conclusion}
We have made the case for using accuracy-efficiency trade-offs as a policymaking lever for assisting in holding distributed ML systems accountable. For AVs, trade-off-informed \emph{ex ante} regulation could constrain the space of undesirable AV behavior, which in turn could narrow the the number of accidents and anomalous behaviors that need to be examined \emph{ex post}. This could lead not only to overall safer behavior, but also the necessary tools to determine accountability when accidents unavoidably occur (Section \ref{sec:policy}). 
More broadly, this discussion can be situated in the context of extracting higher-level values from technical systems---values such as safety and efficiency~\cite{nhtsa2016avs}---as a necessary part of public governance. That is, it is crucial to analyze how higher-level values get implemented via underlying technological mechanisms---in this case, the implementation of the accuracy-efficiency trade-off---to ensure that the implementation aligns with the values that we want to promote in policy. We have argued that the accuracy-efficiency trade-off is not only a correct abstraction, but also the correct level of abstraction, for helping to promote this goal.

Clarifying technical details at this level of abstraction implicates another important value of public governance: transparency. For example, NHTSA has generally does not 
intervene \emph{ex ante} in regulating automobiles \cite{nhtsa2016avs, vinsel2019cars, abraham2019responsibility}. While this might make car 
development more efficient,\footnote{This is a contestable claim. Please refer to \citet{vinsel2019cars} for more details concerning how safety regulations can in fact promote innovations in car technology.} it can come with the 
loss of transparency. Not engaging with technical details \emph{ex ante} can present problems beyond not detecting bugs; it can also lead to not being able to detect whether values like safety 
are implemented appropriately. Worse, 
it is possible that technical values, and the social values they entail, can be deliberately obscured. Technical implementation decisions can be framed as trivial, which can direct policymakers away from viewing them as valid sites for intervention.\footnote{Alternatively, when highly technical jargon is used to describe implementation decisions, it can serve to obfuscate rather than clarify. Rather than enabling transparency for policymakers, who do not tend to be technical experts, these practices can cloud the values at stake~\cite{mulligan2018governance}. In the automotive industry specifically, increasing digital automation has notably led to additional transparency issues, even prior to AVs. Computerized features, in comparison to mechanical ones, can be programmed more easily to obscure true technical performance---for example, to reduce recorded EPA emissions in order to appear more environmentally-friendly \cite{vinsel2019cars}. While out of the scope of this paper, it is worth acknowledging that increased computerization in AVs potentially presents even more transparency issues of this variety.} \citet{mulligan2018governance, mulligan2019ml} have notably written about this issue of technological transparency in public governance. They call out the danger of policy-relevant values decisions getting pushed into 
low-level implementation decisions made by engineers, in place of having the 
values at play being openly debated. This misplacement of responsibility on engineers comes 
from a lack of technical expertise in governance and a resulting lack of 
mechanisms to regulate technology. 
Industry testing and quality control effectively give manufacturers 
the job of converting the law into concrete technical requirements: Manufacturers, instead of public advocacy groups or 
agencies like NHTSA, 
make technical decisions with policy implications without public 
oversight. This conflict-of-interest can lead to compromising or degrading higher-level social values.

We have argued that if policymakers understand the accuracy-efficiency trade-offs in distributed ML systems, and the social values these trade-offs implicate, this problem can (at least in part) be averted. Policymakers will have a more sufficient understanding of technology and will be better able to determine the scope of possibilities for its governance. By understanding the technical values at stake at this level of abstraction, policymakers, with engineers' assistance, could provide insight \emph{ex ante} into how certain implementation decisions should be made. That way, low-level technical matters will not be dismissed as ``just implementation details'' left up to the whims of engineers without public oversight~\cite{mulligan2018governance, jasanoff2016ethics, friedman2019values}. Moreover, when technological failures and accidents do occur---and it is a question of when, not if---rather than viewing them simply as ``unintended consequences'' or ``normal accidents''~\cite{perrow1999risk}, policymakers and other relevant stakeholders 
could more actively participate \emph{ex post} in holding such systems accountable for their behavior. This more-effective public governance 
will improve the power imbalance between system manufacturers and victims of system accidents---empowering and protecting individuals without the resources to seek justice for themselves.


\begin{acks}
This work was made possible by funding from the John D. and Catherine T. MacArthur Foundation and the Digital Life Initiative at Cornell Tech. We thank the Artificial Intelligence Policy \& Practice Initiative at Cornell University and the (Im)perfect Enforcement Conference at Yale Law School for workshopping earlier versions of this work. We also thank the following individuals for their comments and suggestions: Bilan Ali, Jaime Ashander, Harry Auster, Jack Balkin, Solon Barocas, Ken Birman, Fernando Delgado, Thomas G. Dietterich, James Grimmelmann, Steve Hilgartner, David Merritt Johns, Ido Kilovaty, Jon Kleinberg, Kristian Lum, Alan Mackworth, Helen Nissenbaum, Fred B. Schneider, and Matthew Sun.
\end{acks}



\balance
\bibliographystyle{ACM-Reference-Format}
\bibliography{sample-base}



\end{document}